\magnification=1200
\font\text=cmr10
\font\it=cmti10

\font\title=cmbx10 scaled \magstep2

\newcount\notenumber
\def\clearnotenumber{\notenumber=0}
\def\note{\advance\notenumber by 1
\footnote{$^{\the\notenumber}$}}
\clearnotenumber
\baselineskip=18pt
\text
{\baselineskip9pt\vbox{\hbox{FTUV/95-65}}}
{\baselineskip9pt\vbox{\hbox{IFIC/95-68}}}
{\baselineskip9pt\vbox{\hbox{December 1995}}}
\bigskip

\centerline{\title A Flipped SO(10) GUT Model and}
\centerline{\title the Fermion Mass Hierarchy} 

\vskip 1.5 cm

\centerline{{\bf Stefano Ranfone}\footnote{$^\ast$}{E-mail 
RANFONE@evalvx.ific.uv.es} {\bf and Jos\'e W. F. Valle}
\footnote{$^\dagger$}{E-mail VALLE@flamenco.ific.uv.es}}

\medskip

\centerline{\it Instituto de Fisica Corpuscular - C.S.I.C.}
\centerline{\it Department de Fisica Te\`orica, Universitat de Val\`encia}
 \centerline{\it 46100 Burjassot, Val\`encia, SPAIN}

\vskip 3 cm 

\centerline{\bf Abstract}
\vskip 2 cm

\vfill\eject

{\bf Introduction}

Motivated by the nice features of the ``flipped" 
$SU(5)\otimes U(1)$ model [1,2], we have embedded such a model 
in SUSY $SO(10)$. There are many reasons for having 
considered $SO(10)$ [3]. First of all, in this model all 
matter supermultiplets of each generation fit into a 
single irreducible representation; in addition, it includes 
left-right symmetry and yields a more interesting physics 
for neutrino masses. The main motivation for merging the 
idea of ``flipping" with the $SO(10)$ GUT's lies 
in the fact that it gives the possibility of solving 
naturally in these models the problem related to the 
so-called ``doublet-triplet" mass splitting. We recall 
that this problem, common to most of the GUT theories, 
is essentially the need of splitting the mass of the 
standard Higgs doublets with respect to the colour 
triplets, which usually sit in the same supermultiplet. 
The reason for this requirement is that light colour-triplet 
scalars would mediate a too-fast proton decay. On the other 
hand, from the data on $\sin^2 \theta_W\,$ [10] , we also know that 
only one Higgs-doublet pair must remain light, 
developing the {\sl standard} vacuum expectation 
values (VEV's), $v_u$ and $v_d$. One of the best 
solutions of this problem is based on the so-called 
``Missing Partner Mechanism" of Dimopoulos and Wilczek [4]. However, the 
implementation of this idea in realistic GUT models is not 
always so easy, and often requires considerable and unnatural 
modifications [5]. On the contrary, the flipped $SU(5)$ model 
even in its minimal versions has naturally the nice feature of 
pushing up to the GUT scale $M_G$ the mass of the colour-triplets, 
leaving light the standard Higgs doublets [1].

It was early realized that the $SO(10)$ gauge group, among the 
many possible chains of symmetry breaking, could also break 
down to the ``flipped" version of the $SU(5)$ model [6]. 
In particular, this occurs if the {\bf 45}-adjoint Higgs 
supermultiplet, $\Sigma$, develops
\note{$M_{10}$ and $M_5$ represent, respectively, the 
breaking scales of $SO(10)$ down to $SU(5)\otimes U(1)$, 
and of $SU(5)\otimes U(1)$ down to the Standard Model (SM).} 
a large VEV, $M_{10} \geq M_5 \sim 10^{16}$ GeV, along its 
$B$ component, transforming as $(1,0)$ under ``flipped"-
$SU(5)\otimes U(1)$. In the present paper we shall study 
the ``effective" flipped-$SU(5)$ model which is left over 
after the breaking of $SO(10)$, between $M_5$ and $M_{10}$, 
with particular attention to the resulting fermion and Higgs 
mass matrices. In spite of the many advantages of starting 
from $SO(10)$, some difficulties may arise because some 
of the Yukawa-type couplings of the $SU(5)$ superpotential 
have to be equal to each other [7], due to the embedding in $SO(10)$. 
In particular, at a multi-generation level, it is not possible 
to generate any Cabibbo-Kobayashi-Maskawa type of quark mixing, 
unless one enlarges correspondingly also the Higgs doublet sector.
In its turn, this means that also the solution of the above 
mentioned ``doublet-triplet" splitting problem must be modified. 

In this paper, we shall present a self-consistent model 
which solves simultaneously all these difficulties in a 
rather minimal way.  In the last part of the paper, we will
show how this model may be useful for constructing a consistent 
scenario, based on the Georgi-Jarlskog type of texture [8].
We show that, if the first two generations are assumed to get 
their masses via non-renormalizable effective operators,
the model can reproduce the observed quark mass hierarchy 
and mixing, in addition to providing a viable seesaw scheme 
for neutrino masses.

{\bf The Model }

The superfield content of the model is specified in Table 1.
The motivations for our specific choices will be clarified 
during the discussion. For a more detailed description of 
the ``minimal" flipped $SU(5)$ models we refer to the existing 
literature [1,2,7,9].

All standard model fermions plus the right-handed (RH) neutrino, 
$\nu^c$, and their supersymmetric partners, are accommodated into 
three copies of an $SO(10)$-spinorial {\bf 16} superfield:

$$\Psi_i \sim {\bf 16}_i \rightarrow F_i(10,1)\,\oplus 
{\bar f}_i({\bar 5},-3)\,\oplus {l^c}_i(1,5)\,,\eqno(1)$$

\noindent 
where $i=1,2,3$ is the generation index and where 
we have specified its decomposition into $SU(5)\otimes U(1)$. 
The $F_i$, ${\bar f}_i$, and ${l^c}_i$ are the usual flipped 
$SU(5)$ matter superfields, whose particle content may be obtained from
the corresponding assignment of the standard $SU(5)$ GUT model by 
means of the {\sl ``flipping"} $u^{(c)}\leftrightarrow d^{(c)}$, 
$\nu^{(c)}\leftrightarrow e^{(c)}$:

$$\eqalign{\cases{F_i(10,1) = (d,u;d^c;\nu^c)_i\,,\cr
{\bar f}_i({\bar 5},-3) = (u^c;e,\nu)_i\,,\cr
{l^c}_i(1,5) = {e^c}_i\,.\cr}}\eqno(2)$$

\noindent 
The main feature of this ``flipped" assignment is the presence
of $\nu^c$ into the 10-dimensional supermultiplet, while $e^c$ 
is the singlet superfield. The electric charge is specified as 
$Q = T_{3L} - {1 \over 5} Z + {1 \over 5} X$, where $Z$ is the
generator of $SU(5)$ which commutes with $SU(3)_c \otimes SU(2)_L$,
and $X$ is the generator of the extra $U(1)$ symmetry [6].

The first stage of symmetry breaking, $SO(10) \rightarrow SU(5) \otimes U(1)$, 
occurs by assuming that the adjoint-{\bf 45} Higgs supermultiplet 
$\Sigma$, which decomposes under $SU(5)\otimes U(1)$ as:

$$\Sigma \sim {\bf 45}\, \rightarrow S(24,0) \oplus T(10,-4)\oplus \overline{T}
(\overline{10}, 4)\oplus B(1,0)\,,
$$

\noindent gets a VEV along the direction of 
$B(1,0)$, $\,\,<B(1,0)> \equiv M_{10} \geq M_5$.

In the present paper the $SU(5)\otimes U(1) \rightarrow SM$ 
breaking is triggered by two pairs of 16-dimensional 
Higgs supermultiplets, of which  {\sl only one is complete}. 
The other is assumed to be ``incomplete", as in the usual 
minimal-type of flipped models [1,2,7], in order to implement 
the missing-partner mechanism [4].
More explicitly, we assume the presence of the following 
superfields. The ``complete" ${\bf 16} \oplus \overline{{\bf 16}}$ 
pair $\Theta_1 \oplus {\overline \Theta}_1$ decomposes as follows:

$$\Theta_1 \sim {\bf 16}_1 \rightarrow 
\,H_1(10,1)\oplus {\bar \eta}_1({\bar 5},-3)\oplus 
{\xi_1}^c(1,5)\,\,,\eqno(3a)$$

\noindent where:

$$H_1(10,1) = (d_H, u_H; d_H^c; \nu_H^c)_1\,, \,\,\,
{\bar \eta}_1({\bar 5},-3) = (u_H^c; e_H,
 \nu_H)_1\,,\,\,\,\xi_1^c(1,5) = e_{H_1}^c\,;\eqno(3b)$$

\noindent ${\overline \Theta}_1$ of course may be obtained 
from these formulas by just introducing 
the ``bar" over the symbols of the different superfields. 
The ``incomplete" ${\bf 16} \oplus \overline{{\bf 16}}$ 
pair $\Theta_2 \oplus {\overline \Theta}_2$ is, on the 
other hand, identical to the one 
present in the minimal models:

$$\eqalign{\Theta_2 &\sim {\bf 16}_2 \rightarrow 
 \,H_2(10,1)\oplus {\rm ``decoupled\,\,\,\, fields"}\,,
\cr{\overline \Theta_2} & \sim {\overline{\bf 16}}_2 
\rightarrow \,{\bar{H}}_2(\overline{10},-1)\oplus {\rm 
``decoupled\,\,\,\, fields"}\,.\cr}\eqno(4)$$

\noindent 
Their VEV's (assumed to be of order $M_5\sim 10^{16}$ GeV), 
are responsible for the symmetry breaking down to the SM,
$$\eqalign{<\Theta_a> = <H_a>\equiv <\nu_{H_a}^c>
\equiv V_a\,,\,\,\,\,(a=1,2)\,,\cr
<{\overline \Theta}_a> = <{\bar{H}}_a> \equiv <{\bar \nu}_{H_a}^c>
\equiv {\bar{V}}_a\,,\,\,\,\,(a=1,2) \,,\cr}$$

\noindent and are constrained by the D-term flat condition 
as follows: ${V_1}^2 + {V_2}^2 = {{\bar{V}}_1}^2 + {{\bar{V}}_2}^2$.
The presence of the additional GUT-Higgs supermultiplets, 
taken here to be a pair of complete ${\bf 16}\oplus \overline{\bf 16}$
 of $SO(10)$, is one of the 
novelties of the present model. 

The final step of symmetry breaking, down to $U(1)_{em}$, 
is due to the VEV's developed by two different 10-dimensional 
$SO(10)$ superfields:

$$\Delta_{\alpha} \sim {\bf 10}_\alpha \rightarrow 
h_{\alpha}(5,-2)\oplus {\bar h}_{\alpha}({\bar 5},2)\,,\eqno(5)$$

\noindent $(\alpha = 1,2)$, in strict analogy with the 
standard type of $SO(10)$ GUT models. The multiplets
$h_{\alpha}$ and ${\bar h}_{\alpha}$ contain the SM 
Higgs doublets plus colour-triplets and anti-triplets, 
respectively:

$$h_{\alpha}(5,-2) = (D; h^-, h^o)_{\alpha}
\,,\,\,\,\,\,\,\,{\bar h}_{\alpha}({\bar 5},2) = 
({\bar D};{\bar h}^o, {\bar h}^+)_{\alpha}\,.\eqno(6)$$

\noindent 
Notice the different $U(1)$ quantum numbers of these 5-dimensional 
representations with respect to ${\bar f}_i$ of the matter 
supermultiplets.

The last ingredient of our model is an $SO(10)$ singlet superfield 
$\Phi\sim {\bf 1} \rightarrow \phi(1,0)$, whose VEV, $\sigma$, 
will be used for generating the observed hierarchical pattern 
of fermion masses, as we discuss in detail in the last
part of the paper.

In order to further specify our model we start writing the 
general renormalizable cubic superpotential as follows:

$$\eqalign{{\cal W}_{10}^R =& A_\alpha^{ij} \Psi_i 
\Delta_\alpha \Psi_j + B_\alpha^{ab} \Theta_a 
\Delta_\alpha \Theta_b + C_\alpha^{ab} {\overline \Theta}_a 
\Delta_\alpha {\overline \Theta}_b + D^{i,a} \Psi_i {\overline \Theta}_a 
\Phi \cr &+
 E_{\alpha \beta} \Delta_\alpha \Delta_\beta \Phi + F_\alpha^{i,a} \Psi_i \Theta_a 
\Delta_\alpha + G \Phi^3 + I_{a b} \Theta_a {\overline \Theta}_b \Phi + L_{\alpha \beta} 
\Sigma \Delta_\alpha \Delta_\beta\,\,,\cr}\eqno(7)$$

\noindent where $A, B, C, E, L$ are symmetric matrices,
$i,j = 1,2,3$ label the different 
matter generations, $a,b = 1,2$ (``1" and ``2" representing 
respectively the ``complete" and the ``incomplete" spinorial 
superfields), and $\alpha,\, \beta = 1,2$ label the two 
different 10-dimensional Higgs supermultiplets. As we will
show later,  phenomenology and the requirement of a 
{\sl correct} doublet-triplet mass-splitting will constrain 
the structure of the above superpotential.  Under the 
$SO(10) \rightarrow SU(5)\otimes U(1)$ breaking, 
induced by $< \Sigma >$, we have:

$$\eqalign{{\cal W}_{10}^R & \rightarrow {\cal W}_5^R =  
A_\alpha^{ij} (F_i F_j h_\alpha + F_i {\bar f}_j
{\bar h}_\alpha + {\bar f}_i l_j^c h_\alpha ) + 
B_\alpha^{ab} ( H_a H_b h_\alpha + H_a 
{\bar \eta}_b {\bar h}_\alpha \delta_{b1} \cr & + 
{\bar \eta}_a \xi_b^c h_\alpha \delta_{a1} \delta_{b1}) +
C_\alpha^{ab} ( {\bar H}_a {\bar H}_b 
{\bar h}_\alpha + {\bar H}_a \eta_b 
h_\alpha \delta_{b1} + \eta_a {\bar \xi}_b^c {\bar h}_\alpha \delta_{a1} 
\delta_{b1}) + D^{i,a} ( F_i {\bar H}_a \cr 
& + {\bar f}_i \eta_a \delta_{a1} + l_i^c {\bar \xi}_a^c ) \phi +
E_{\alpha \beta} h_\alpha {\bar h}_\beta \phi + F_\alpha^{i,a} 
( F_i H_a h_\alpha + {\bar f}_i H_a {\bar h}_\alpha + 
F_i {\bar \eta}_a {\bar h}_\alpha \delta_{a1} \cr & +
{\bar f}_i \xi_a^c h_\alpha \delta_ {a1} + l_i^c {\bar \eta}_a 
h_\alpha \delta_{a1} ) +
G \phi^3  + I_{a b} ( H_a {\bar H}_b + {\bar \eta}_a \eta_b 
\delta_{a1} \delta_{b1} + \xi_a^c {\bar \xi}_b^c \delta_{a1} 
\delta_{b1} )\phi \cr &+ L_{\alpha \beta} ( T {\bar h}_\alpha 
{\bar h}_\beta + {\bar{T}} h_\alpha h_\beta + B h_\alpha 
{\bar h}_\beta + S h_\alpha 
{\bar h}_\beta )\,,\cr}\eqno(8)$$

\noindent where $\delta_{ab}$ is the Kronecker symbol.

Since the particle content of the present model is quite different with 
respect to the minimal versions already studied in the literature, 
it may be worthwhile to list the Goldstone bosons eaten by the 
gauge bosons and the left-over (physical) fields which will 
enter in the mass matrices. In the symmetry breaking
$SU(5)\otimes U(1) \rightarrow SM$ the Goldstone bosons 
``absorbed" by the $X,\,Y$ gauge bosons\note{
Similarly, their conjugate fields $\bar{X}$ and $\bar{Y}$ will absorb a 
corresponding combination of
$({\bar d}_{H_1}$, ${\bar u}_{H_1}) \in {\bar H}_1({\overline{10}},-1) 
\subset {\bar\Theta}_1$ and $({\bar d}_{H_2},,\,{\bar u}_{H_2}) \in 
{\bar H}_2({\overline{10}},-1) \subset {\bar\Theta}_2$.}
are a linear combination of $(d_{H_1}\,,\,u_{H_1}) \in H_1(10,1) \subset 
\Theta_1$ and $(d_{H_2}\,,\,u_{H_2}) \in H_2(10,1) \subset \Theta_2$. 
Their orthogonal combinations remain ``uneaten" 
and will mix, in general, with the standard {\sl up-} and 
{\sl down-}type squarks. We shall denote these fields simply 
by $(d_H, \,u_H)$ and $({\bar d}_H,\, {\bar u}_H)$. On the
other hand, the heavy neutral gauge boson gets its large mass 
by absorbing a linear combination of $\nu^c_{H_1}$, $\nu^c_{H_2}$,
${\bar\nu}^c_{H_1}$ and ${\bar\nu}^c_{H_2}$.

At this point we may study the ``doublet-triplet" splitting 
problem, starting from the $SU(2)_L$-{\sl doublet} 
mass matrix ${\cal M}_2$. The model 
contains six ``down-type" doublets
\note{For notational simplicity, sometimes we will represent 
the ``doublet" part by the same symbol used for the 
corresponding {\bf 5}-plet ({\it e.g.}, ${\bar f}$, ${\bar h}$, etc.).}:
${\bar f}_i \equiv (e,\nu)_i \subset \Psi_i\,,\,\,(i=1,2,3)$;  
$\,\,\,h_\alpha\equiv (h_\alpha^-, h_\alpha^o) \subset \Delta_\alpha\,,\,\,(\alpha=1,2)$; and ${\bar \eta}\equiv (e_H, \nu_H) \subset \Theta_1$. 
On the other hand, there are only three ``up-type" doublets:
${\bar h}_\alpha\equiv ({\bar h}^o_\alpha , 
{\bar h}^+_\alpha) \subset \Delta_\alpha\,,\,\,
(\alpha =1,2)$, and $\eta\equiv ({\bar\nu}_H, 
{\bar e}_H)\subset {\bar{\Theta}}_1$. 
The mass terms for these fields generated by the superpotential 
${\cal W}_5^R$ yield the following $3 \times 6$ doublet mass matrix:

$${\cal M}_2 \,=\, \bordermatrix{&{\bar f}_i&h_\alpha &{\bar\eta}\cr
{\bar h}_\beta&F_\beta^{i,a} V_a& E_{\alpha\beta} \sigma+
L_{\alpha\beta}M_{10}&B_\beta^{a1} V_a\cr
\eta&D^{i,1}\sigma&C_\alpha^{a1} {\bar{V}}_a&I_{11}\sigma\cr}\,.\eqno(9)$$

\noindent 
The phenomenological constraints on the entries of this matrix are 
the following. First of all, the data on $\sin^2\theta_W$ [10] require that 
only one Higgs doublet pair must remain light (at the weak scale),
in order to develop the two electroweak VEV's, $v_u$ and $v_d$ 
(such that $v_u^2+v_d^2= v_{SM}^2 \simeq (246\,\, {\rm GeV})^2\,$). 
All other pairs are, on the other hand, required to be very heavy, 
so as to prevent them from acquiring a non-zero VEV. This means that 
${\cal M}_2$ must be a {\sl rank} = 2 matrix. Finally, it is desirable 
to avoid any mixing ($\propto M_G$) between ``matter" and ``Higgs" 
superfields. All these requirements may be satisfied by imposing: 

$$F_\alpha^{i,a} = D^{i,a} = 
E_{\alpha\beta} = L_{\alpha\beta} = 0\,,\eqno(10)$$

\noindent in the superpotential (7,8).
\noindent In this case, leaving aside the massless matter
doublet superfields (${\bar f}_i$) which decouple from the other 
doublets, we get a {\sl reduced} matrix ${\tilde{\cal M}}_2$:

$${\tilde{\cal M}}_2 \,= \, \bordermatrix{&h_1&h_2&{\bar\eta}\cr
{\bar h}_1&0&0&B_1\cr {\bar h}_2&0&0&B_2\cr \eta&C_1&C_2&I_1\cr}\,,\eqno(11)$$

\noindent where we have set:

$$B_a = \sum_{b=1}^2 B_a^{b1} V_b\,,\,\,\,C_a = \sum_{b=1}^2 C_a^{b1} {\bar V}_b\,,\,\,(a=1,2)\,,\,\,\,
{\rm and} \,\,\,\, I_1\equiv I_{11} \sigma\,.\eqno(12)$$

\noindent 
Clearly, ${\tilde{\cal M}}_2$ is a singular rank-2 matrix.
The corresponding (hermitian) squared mass matrix 
${\tilde{\cal M}}_2^\dagger {\tilde{\cal M}}_2$ will yield 
only two non-zero eigenmasses, for the two heavy doublet pairs:

$$m_{2,3}^2 = {1\over 2} \left(\Sigma^2 + I_1^2\right) \pm {1\over 2} \sqrt{\Pi^4+ 2 \Sigma^2 I_1^2 + 
I_1^4}\,,$$

\noindent where:  $\Sigma^2 = {B_1}^2 + {B_2}^2 + {C_1}^2  + {C_2}^2\,$, and 
$\Pi^2 = {B_1}^2 + {B_2}^2 - {C_1}^2  - {C_2}^2\,$. Of course, the 
other eigenvalue of ${\tilde{\cal M}}_2^\dagger {\tilde{\cal M}}_2$ 
is zero, corresponding to the MSSM single pair of light Higgs doublets.
 
As we shall discuss in the last part of the paper, we ascribe 
the observed hierarchy among the fermion masses to the existence 
of non-renormalizable terms in the superpotential, proportional 
to powers of the ``suppression factor" 
$<\Phi>/M_P\,$
($M_P$ being the Planck mass).
The singlet field VEV $\sigma \equiv <\Phi>$ can then be fixed 
by using the measured value of the Cabibbo angle 
\note{We recall that, within specific ``texture" models, 
$\sin\theta_C \sim \sqrt{m_1/m_2}\,$, where $m_1$ and $m_2$ are, 
respectively, the masses of the first and the second generation 
(either {\sl up-} or {\sl down-}) quarks.}:

$$\lambda\equiv {\sigma\over M_P}\sim \sin \theta_C \simeq 0.22\,,\eqno(13)$$

\noindent 
which gives $\sigma \sim 10^{18}$ GeV. This value is 
sufficiently larger than the GUT scale, $M_5$, so that 
we can assume $I_1 \gg B_i\,, \,\,C_i\,$. As a result we
can expand the formulas for the eigenmasses of the heavy Higgs 
doublet pairs, and get approximate expressions for the 
corresponding eigenvectors. A simple calculation gives:

$$m_1 = 0 \,,\,\,\,\,m_2\simeq{1\over 2 I_1}
\sqrt{\Sigma^4-\Pi^4}= {1\over I_1}\sqrt{({B_1}^2 + {B_2}^2) \,
({C_1}^2 + {C_2}^2)}\,,\,\,\,\,m_3\simeq I_1\,,\eqno(14)$$

\noindent 
which shows that the two massive doublet pairs have masses
of order $m_2\sim {M_5}^2/\sigma\sim{\cal O}(10^{14} \,{\rm GeV})$ 
and $m_3\sim\sigma\sim{\cal O}(10^{18}\, {\rm GeV})$, respectively.
Then, we may also get the expressions for all the 
(left and right) eigenvectors of ${\tilde{\cal M}}_2$. 
In particular, we find that the heaviest fields
correspond to the $\eta$ and ${\bar\eta}$ coming 
from the ``complete" GUT-Higgs 
${\bf 16} \oplus {\bf{\overline{16}}}$ and have
a mass $m_3\sim \sigma$. Moreover, they do not mix, 
up to terms of order $\lambda$, 
with the other doublets. On the other hand, the MSSM 
Higgs bosons (with $m_1=0$) and the heavy doublets 
with a mass $m_2\sim {M_5}^2/\sigma$, correspond  
to two orthogonal combinations of the fields $h_\alpha$ 
and ${\bar h}_\alpha$. Denoting these mass eigenstates as 
$H_{u(d)}$ and $\Phi_{u(d)}$,  respectively, we find:

$$\eqalign{H_u &= {\bar h}_1 \cos\chi - {\bar h}_2 \sin\chi\,,\cr
\Phi_u &= {\bar h}_1 \sin\chi + {\bar h}_2 \cos\chi\,,\cr}\eqno(15a)$$

\noindent and, analogously:

$$\eqalign{H_d &= h_1 \cos\xi - h_2 \sin\xi\,,\cr
\Phi_d &= h_1 \sin\xi + h_2 \cos\xi\,,\cr}\eqno(15b)$$

\noindent where the two mixing angles are given by:

$$\tan \chi = {B_1\over B_2}\,,\,\,\,\,\,\,\tan \xi = 
{C_1\over C_2}\,.\eqno(16)$$

\noindent The ``doublet" mass Lagrangian will then be written as:

$${\cal L}_2 = m_1 H_u H_d + m_2 \Phi_u \Phi_d + m_3 \eta {\bar\eta}\,.$$

\noindent 
This concludes our study of the doublet sector.

Let's turn now our attention to the (colour) ``triplets". 
The fields left over after the GUT symmetry breaking 
transforming as ${\bf 3}$ under $SU(3)_c$ are the following:
$(u_i,d_i)\in F_i(10,1) \subset \Psi_i \,\,\,(i=1,2,3)$, coming 
from the ordinary matter supermultiplets, $(u_H, d_H)$ 
(which is the linear combination, coming from 
$H_1(10,1)\oplus H_2(10,1)\subset 
\Theta_1\oplus\Theta_2$, orthogonal
to the Goldstone bosons eaten by the $(X,Y)$ gauge bosons, 
as we have mentioned above), 
${\bar u}_H^c \in \eta_1(5,3) \subset {\bar\Theta}_1$, 
${\bar d}_{H_a}^c \in {\bar H}_a({\overline 10},-1) 
\subset {\bar\Theta}_a\,,\,\,(a=1,2)$, and
$D_\alpha\in h_\alpha(5,-2)\subset\Delta_\alpha\,, \,\,(\alpha =1,2)$ 
(from the electroweak Higgs 
superfields). Of course, we also have the corresponding antiparticles 
transforming as ${\bar{\bf 3}}$ under $SU(3)_c$. Taking into account 
the conditions given in eq.(10) and using the fact that $\eta$
and ${\bar \eta}$, being very heavy fields, cannot develop a 
non-zero VEV, we may easily get from the superpotential the 
mass matrices for the {\sl up}- and the {\sl down}-type 
colour-triplets. In particular, we notice that the matter 
superfields ($d_i\,,\,\,\, d_i^c\,,\,\,\, u_i\,,\,\,\,u_i^c\,$) do 
not receive any contribution proportional to the GUT scale
and do not mix with the other states, so that they remain 
exactly massless before the electroweak symmetry breaking. 
This fact allows us to express these mass matrices simply 
in the following form:

$${\cal M}_{3,d} \,= \, \bordermatrix{&{\bar d}_H&{\bar D}_\beta&d^c_{H_b}\cr
d_H&I\sigma&0&0 \cr D_\alpha&0&0&B_\alpha^{a b} V_a \cr {\bar d}^c_{H_a} & 0 & C_\alpha^{a b}
{\bar{V}}_a&I_{a b}\sigma\cr},\,\,\,\,(a,b,\alpha,\beta=1,2),\eqno(17a)$$

\noindent and

$${\cal M}_{3,u} \,= \, \bordermatrix{&{\bar u}_H&u^c_H\cr
u_H&I \sigma&0\cr {\bar u}^c_H&0&I_{11} \sigma\cr}\,,\eqno(17b)$$

\noindent where $I$ stands for some combination of the coefficients $I_{a b}$.
Let us analyse these matrices. First of all, we see that the 
``uneaten" combinations of the Higgs superfields,  
$d_H, u_H$, and their antiparticles, get 
a very large mass proportional to $\sigma\sim
10^{18}\,$ GeV. In addition, we see that in general 
both the $5\times 5$ matrix $\,{\cal M}_{3,d}\,$ and the 
$2\times 2$ matrix $\,{\cal M}_{3,u}\,$ will be non-singular,
ensuring that all the corresponding triplet fields will get a mass 
at the GUT scale. More precisely, in the ``down" sector,  
three states will get a mass of order $\sigma$, while the 
other two will have a mass proportional to
$V{\bar{V}}/\sigma\sim M_G^2/\sigma$. On the other hand, 
${\cal M}_{3,u}$ suggests that both the
``up"-type fields will have a mass of order $\sim\sigma$. 
The heaviness of all these colour triplets
prevents the presence of unsuppressed dimension five
operators [20] which might mediate a too fast proton decay. 
The fact that we have been able to get a single light 
Higgs doublet pair, while ensuring the absence of 
dangerous light colour triplets, is essentially 
the solution of the doublet-triplet splitting 
problem, for our ``extended" flipped model.

Let us now turn our attention to the Yukawa couplings 
responsible for generating the observed masses 
for the standard model fermions. All fermion masses 
arise from the $A$-term in the superpotential
given in eqs.(7-8). Rewriting the ``interaction" 
eigenstates $h_\alpha\,$ and $\,{\bar h}_\alpha\,$ 
in terms of the ``physical" states 
$H_{u(d)}\,$ and $\,\Phi_{u(d)}\,$ by means of 
eqs.(15), we may write the Yukawa Lagrangian as:

$$\eqalign{{\cal L}_Y \,&=\, ({\bar e}_{Li} e_{Rj} + {\bar d}_{Li} d_{Rj}) \left[ A_1^{ij} (H_d 
\cos\xi +
\Phi_d \sin\xi) + A_2^{ij} (- H_d \sin\xi + \Phi_d \cos\xi)\right] \cr &+
[6~({\bar u}_{Li} u_{Rj} + {\bar \nu}_{Li} \nu_{Rj}) \left[ A_1^{ij} (H_u \cos\chi +
\Phi_u \sin\chi) + A_2^{ij} (- H_u \sin\chi + \Phi_u \cos\chi)\right]\,. \cr\,}\eqno(18)$$

\noindent 
Since in the electroweak symmetry breaking only the light fields 
$H_u$ and $H_d$ may develop non-zero VEVS, $v_u$ and $v_d$, 
respectively, from eq.(18) we get the following mass matrices:

$$\eqalign{M_d^{ij} = M_e^{ij} &= (A_1^{ij} 
	\cos\xi - A_2^{ij} \sin\xi ) v_d\,,\cr
	M_u^{ij} = M_{\nu D}^{ij} &= (A_1^{ij} 
	\cos\chi - A_2^{ij} \sin\chi ) v_u\,,\cr}\eqno(19)$$

\noindent 
where $A_{1(2)}$ are the $3\times 3$ Yukawa coupling matrices, 
the mixing angles $\xi$ and $\chi$ were given in terms of the 
parameters of the superpotential through eqs.(16) and
(12), and as usual, $v_u/v_d \equiv \tan \beta$.

Eq.(19) shows that, since in general $\xi \neq \chi$, the quark 
mass matrices $M_u$ and $M_d$ will {\sl not} be proportional 
to each other, resulting in a non-trivial KM mixing. We recall 
that this fact, in the context of our flipped $SO(10)$  model, 
is the result of enlarging the electroweak Higgs sector
to two {\bf 10}-plets. The main result of the work carried 
out in this first part of the paper has been the extension of 
the ``missing partner" solution of the doublet-triplet 
splitting problem to this non-minimal model. This 
extension  has been obtained by introducing a ``complete" 
(second) pair of ${\bf 16} \oplus {\overline{\bf 16}}$ 
GUT Higgs superfields. We have also taken into account the 
constraint, derived from the data on $\sin^2\theta_W$, of 
having only one light electroweak Higgs doublet pair.

\noindent {\bf Non-Renormalizable terms, Textures, and 
the quark mass hierarchy} 

One of the more successful {\sl ansatze} for the up- and 
the down-quark mass matrices is the one proposed several 
years ago by Georgi and Jarlskog [8], according to which:

$$M_u = \left(\matrix{0&a&0\cr a&0&b\cr 0&b&c\cr}\right)\,,\,\,\,\,
 M_d = \left(\matrix{0&d e^{i\phi}&0\cr d e^{-i\phi}
& f & 0 \cr  0 & 0 & g\cr}\right)\,
 M_e = \left(\matrix{0 & d e^{i\phi} & 0 \cr d 
e^{-i\phi} & -3 f & 0 \cr 0 & 0 & g \cr}\right)\,,\eqno(20)$$

\noindent 
where all parameters $a,b,c,d,f,g,\phi$ are taken to be real.
We recall that in the Georgi-Jarlskog scheme
it is possible to fit the mass for all generations 
of charged leptons, in view of the $-3$ factor which 
multiplies the parameter $f$ in the (22)-element of 
the matrix $M_e$. 
In the framework of the standard $SU(5)$ GUT models, 
such $-3$ factor may be obtained via the contribution 
of a 45-dimensional Higgs supermultiplet (coming from the ${\bf 126}$
of $SO(10)\,$). However, such a large representation is not 
required for breaking the ``flipped" $SU(5)$ gauge group [1,2], 
and is not even allowed if we derive the model from the
superstring (at a Kac-Moody algebra level equal to one). 

Here we shall not try to explain the masses of the 
first two generations of charged leptons. We simply 
keep the simple mass relation of minimal GUT models, 
$m_{e_i}=m_{d_i}\,$ (at the GUT scale), which is 
very successful for the third generation, leading 
to the prediction $m_b \simeq r m_\tau$ at low-energy, 
$r (\simeq 2.7)$ being the appropriate renormalization 
parameter. In other words, in this paper we choose not 
to use the full Georgi-Jarlskog ansatz including the 
charged leptons. For further simplicity we shall also 
assume all mass matrices to be real, thus neglecting
CP violation. 

A way for understanding the observed fermion mass 
hierarchy in the Georgi-Jarlskog scheme is to assume a 
corresponding hierarchical pattern for the parameters 
of the mass matrices: $\,a\ll b\ll c\,$ and $\,d\ll f\ll g\,$. 
In this case one can easily check that the quark masses 
at the GUT scale ({\it i.e.}, the eigenvalues of $M_u$ and $M_d$) 
are given by the following formulas:

$$m_t\simeq c\,,\,\,\,\, m_c\simeq {b^2\over c}\,,\,\,\,\, 
  m_u\simeq \left({a\over b}\right)^2 c\,, 
\,\,\,\,m_b\simeq g\,,\,\,\,\, m_s\simeq f\,,\,\,\,\,m_d\simeq 
{d^2\over f}\,.\eqno(21)$$

\noindent 
The corresponding diagonalizing matrices for the {\sl up} and the 
{\sl down} quark sectors can be expressed in terms of three
mixing angles, $\theta_i\,$, as follows:

 $$\eqalign{V_u\,=&\,\left(\matrix{c_2&s_2&0\cr -s_2&c_2&0\cr 0&0&1\cr}\right)\,\,\,
 \left(\matrix{1&0&0\cr 0&c_3&s_3\cr 0&-s_3&c_3\cr}\right)\,,\cr
V_d\,=&\,\left(\matrix{c_1&s_1&0\cr -s_1&c_1&0\cr 0&0&1\cr}\right)\,\,\,
,\cr}\eqno(22)$$ 

\noindent 
where we have set $s_i \equiv \sin\theta_i\,$ and 
$c_i \equiv \cos\theta_i\,\,(i=1, \dots , 3)$. These
are given approximately by:

$$s_1 \simeq \sqrt{m_d\over m_s}\simeq {d\over f}\,,\,\,\,\,
  s_2 \simeq \sqrt{m_u\over m_c}\simeq {a c\over b^2}\,,\,\,\,\,
  s_3 \simeq \sqrt{m_c\over m_t}\simeq {b\over c}\,\,\,\,\,.\eqno(23)$$

\noindent  
The elements of the ordinary KM mixing matrix $V_{KM}$, 
which is just the product $V_u^\dagger\,V_d$, may then be 
written as: $|V_{us}| \simeq |V_{cd}| \simeq \sin\theta_C \simeq s_1\,,$
$|V_{cb}|\simeq s_3\,,$ and $|V_{ub}/V_{cb}|\simeq s_2$. 
Renormalizing the quark masses from $M_G$ down to low-energy 
it is found that these formulas give a sufficiently good 
agreement with the present experimental data on quark mixing
[11]: 
$|V_{us}| \simeq |V_{cd}|\simeq 0.221 \pm 0.003$, 
$\,|V_{cb}| \simeq 0.040 \pm 0.008$, and $\,|V_{ub}| \simeq 
0.003 \pm 0.002$. 
This means that in the framework of the Georgi-Jarlskog ansatz, 
the understanding of the quark mixing pattern is reduced to that 
of the hierarchical structure of the quark mass spectrum.

A standard strategy for facing the problem related to the explanation 
of the fermion mass hierarchy in the context of GUT's is to assume 
that the mass is generated via the standard Higgs mechanism only 
for the third family fermions, while the first two generations 
get their mass only through higher-dimensional non-renormalizable 
operators in the superpotential. In the present model 
we shall employ such non-renormalizable operators also for producing an 
``effective" {\sl seesaw} mechanism, needed for suppressing the 
neutrino masses.

The simplest possibility is to assume that all the hierarchy is 
due to powers of a single ``suppression" factor, $\lambda$, equal 
to the ratio of the VEV of the singlet field, $<\Phi> \equiv \sigma$, 
and the Planck mass, $M_P$, which may be considered as the ultimate 
cut-off of the theory. 

In what follows it is important to notice that all quark mass ratios,
normalised at the same scale $M_G$, can be expressed in terms of
the universal parameter $\lambda \sim \sin \theta_C \simeq 0.22$
({\it i.e.}, the ``$\lambda$-parameter" of the Wolfenstein 
parametrisation for the quark mixing matrix),
according to the following pattern:

$$\eqalign{{m_u\over m_t} &\simeq \left({a\over b}\right)^2\sim \lambda^8\,,\,\,\,\,\,\,
           {m_d\over m_b}  \simeq \left({d^2\over f g}\right)\sim \lambda^4\,,\cr
           {m_c\over m_t} &\simeq \left({b\over c}\right)^2\sim \lambda^4\,,\,\,\,\,\,\,
           {m_s\over m_b}  \simeq \left({f\over g}\right)\sim \lambda^2\,,\cr}\eqno(24)$$

\noindent 
These eqs. yield the following relations:

$$b \simeq \lambda^2  c ,\,\,\, a \simeq \lambda^6 c ,\,\,\,
f\simeq \lambda^2 g ,\,\,\, d \simeq \lambda^3 g\,,\eqno(25)$$

\noindent from which we can write the two quark mass matrices as follows:

$$M_u \simeq \left(\matrix{0&\lambda^6&0\cr 
\lambda^6&0&\lambda^2\cr 0&\lambda^2&1\cr}\right)\,m_t\,,\,\,\,\,\,
M_d \simeq \left(\matrix{0&\lambda^3&0\cr 
\lambda^3&\lambda^2&0\cr 0&0&1\cr}\right)\,m_b\,.\eqno(26)$$

\noindent 
As we see, only the third generation quarks get their mass 
via the standard Higgs mechanism, implemented in the renormalizable 
superpotential given in eqs.(7-8). 

From eqs.(23-25) we easily get:

$${<\Phi>\over M_P}\equiv \lambda\sim\sqrt{m_d\over m_s}
\sim\sin\theta_C\simeq 0.22\,;\eqno(27)$$

\noindent 
The fact that the suppression factor $\lambda$ is approximately equal 
to the Cabibbo angle fixes the VEV of our singlet superfield, 
$<\Phi> \equiv \sigma \simeq \,0.22\, 
M_P\sim {\cal O}(10^{18})\,$ GeV, very near the Planck scale, 
and much larger than the GUT scale, $M_G$, thus justifying
the expansion with respect to the ratio $M_G/\sigma$ we used
in eq.(14).

At this point we wish to show how it is possible to get the 
mass matrices given in eqs.(26) in the framework of the 
present model. From the comparison of eqs.(19) and (26) 
we get for the two ($3\times 3$) Yukawa-coupling matrices 
$A_1$ and $A_2$ the following expressions:

$$\left(\matrix{A_1\cr A_2\cr}\right)\,=
\,{\cal D}^{-1}\,\left(\matrix{-\sin\chi &\sin\xi\cr
-\cos\chi &\cos\xi\cr}\right)\,\left(\matrix{\mu_d\cr \mu_u\cr}\right)\,,
\eqno(28)$$

\noindent 
where $\,\,{\cal D} \equiv \sin (\xi - \chi) \,$ 
and $\mu_{u(d)}\equiv M_{u(d)}/v_{u(d)}$. 
Writing explicitly the matrix elements we find:

$$\eqalign{A_1^{33} &= {\cal D}^{-1}\,
\left(-y_b\sin\chi+y_t\sin\xi\right)\,,\cr
A_1^{23} &= A_1^{32} =  {\cal D}^{-1}\,y_t \lambda^2 \sin\xi\,,\cr
A_1^{22} &=  -{\cal D}^{-1}\,y_b \lambda^2 \sin\chi\,,\cr
A_1^{13} &=A_1^{31}=0\cr
A_1^{12} &=A_1^{21} = {\cal D}^{-1}\,
\left(-y_b\lambda^3\sin\chi +y_t\lambda^6\sin\xi\right)\,,\cr
A_1^{11} & = 0 \,,\cr}\eqno(29a)$$

\noindent and

$$A_2^{ij} = A_1^{ij}(\sin\xi\rightarrow\cos\xi\,,\,\,
\sin\chi\rightarrow\cos\chi)\,,\eqno(29b)$$

\noindent 
where $y_b\equiv m_b/v_d\,$ and $y_t\equiv m_t/v_u$ are
the bottom and top quark Yukawa couplings. These formulas completely fix 
the structure we need to assume for the non-renormalizable 
(NR) terms in the superpotential, required in order to reproduce 
the correct quark mass spectrum and mixing.

The part of the superpotential responsible for generating 
fermion masses (apart from the term needed for implementing 
the neutrino seesaw mechanism, which will be discussed later) may 
then be expressed as a series expansion in powers of the 
suppression factor $\lambda$ as follows:

$${\cal W}^{10}_Y\, = \,\underbrace{A_{(R),\alpha}^{ij}\,
\Psi_i\Delta_\alpha\Psi_j}_{\rm\, renormal.} +
\underbrace{\sum_n \,A_{(n),\alpha}^{ij}\,\left({\Phi\over M_P}\right)^n\,
\Psi_i\Delta_\alpha \Psi_j}_{\rm\, non-renormal.}\, . \eqno(30)$$

\noindent 
In the above equation 

$$A_{(R),\alpha} = 
\left(\matrix{0&0&0\cr0&0&0\cr0&0&A_\alpha^{33}\cr}\right)\,,\,\,\,\,
(\alpha=1,2)\,,\eqno(31)$$

\noindent 
($A_1^{33}\,$ and $\,A_2^{33}\,$ have been given in eqs.(29)), 
showing that at the renormalizable level only the third generation 
fermions get mass. For the non-renormalizable 
part,  eqs.(29) give the following non-vanishing
Yukawa-type couplings $\,A_{(n),\alpha}^{ij}\,$:

$$\eqalign{A_{(2),1}\,& = \, {\cal D}^{-1}\, 
\left(\matrix{0&0&0\cr 0&-y_b\sin\chi&y_t\sin\xi\cr
0 & y_t\sin\xi&0\cr}\right)\,,\cr
A_{(3),1} \,& = \, {\cal D}^{-1}\, 
\left(\matrix{0&-y_b\sin\chi&0\cr -y_b\sin\chi&0&0\cr0&0&0\cr}\right)
\,,\cr
A_{(6),1}\,&= \, {\cal D}^{-1}\, 
\left(\matrix{0&y_t\sin\xi&0\cr y_t\sin\xi&0&0\cr0&0&0\cr}\right) 
\,,\cr}\eqno(32a)$$

\noindent and:

$$A_{(n),2} = A_{(n),1}(\sin\chi\rightarrow\cos\chi\,,\,\,\,\,\sin\xi\rightarrow\cos\xi)\,.\eqno(32b)$$

\noindent
These couplings fix completely the superpotential in eq.(30), 
responsible for the generation of the charged fermion mass 
hierarchy.  

We now turn to the neutrino masses. 
As we have seen above in eq.(19), the ``Dirac" neutrino 
mass matrix $M_{\nu D}$ is expected to be equal to the 
up-quark mass matrix $M_u$ at the GUT scale. This means, 
of course, that we need to implement a seesaw-type of mechanism, 
in order to suppress the neutrino masses consistently with the present
phenomenological (astrophysical and cosmological) bounds [12].

As is well known, a key ingredient of the seesaw mechanism [13],
at least in the framework of the ordinary GUT's, is the presence
of a large Majorana mass for the right-handed (RH) neutrinos ($\nu^c$). 
In the standard $SO(10)$ models, for example, this mass can be 
generated via the contribution of the $SU(2)_L$-singlet Higgs field
sitting in the {\bf 126}-representation. In the non-supersymmetric 
models, it may also be induced radiatively at the two-loop level 
via the so-called Witten mechanism [14], even in the absence of the 
{\bf 126} Higgs multiplet. In such a case, it is sufficient to 
employ just the VEV at the GUT scale of a {\bf 16} Higgs 
multiplet, also responsible  for breaking the $SO(10)$ 
gauge group down to $SU(5)$.
On the other hand, in flipped string-inspired 
models based on $SU(5)\otimes U(1)$, due to the 
lack of large Higgs representations, it is
not possible to generate a Majorana mass for the 
RH neutrino. However, an effective seesaw mechanism may be 
produced by means of the mixing at the GUT scale 
between the $\nu^c$ and the fermionic component 
of extra singlet superfields, $\phi_i$ [15]
In these minimal type models, this mixing was 
produced by a term of the type 
$F_i\,<{\bar H}>\,\phi_j\rightarrow \nu^c_i\,\phi_j\,M_G\,$. 
Unfortunately,  such a term disappears in our extended 
($SO(10)$-embedded) model, since, as can be seen from 
eq.(8), it is proportional to the Yukawa-type couplings $D^{i,a}$, 
which had to be set to zero according to the conditions (10). Therefore,
we need to construct a seesaw type of model in a different way. 
Consistently with the philosophy adopted in our study of the 
charged fermion mass hierarchy we shall assume that also the 
seesaw mechanism is due to NR terms in the superpotential [16].
A minimal choice, as was discussed in ref. [7], 
is the following:

$${\cal W}_{\nu^c}\,=\,{1\over M_P}\, \Gamma_{ab}^{ij}\, \Psi_i 
{\bar \Theta}_a {\bar \Theta}_b \Psi_j
\,,\,\,\,\,\,\,(a,b=1,2;\,\,\,i,j=1,2,3).\eqno(33)$$
     
\noindent 
Since only the ${\bf 10}$-components (${\bar H}_a$) of ${\bar\Theta}_a$ 
may develop a non-vanishing 
\note{In fact, as we have seen from eq.(11), the {\bf 5}-component of 
${\bar\Theta}_1\,$, $\eta\,$, having a very large mass 
$m_3\simeq I_1\sim {\cal O}(\sigma)$, cannot develop a non-zero VEV.}  
VEV ($\,<{\bar H}_a>\equiv <{\bar\nu}^c_{H_a}> \equiv  
{\bar V}_a\,,\,\,\,(a=1,2)\,$ ), it is easy to see that 
the term ${\cal W}_{\nu^c}$ may affect only the RH neutrino 
component of the matter superfields $\Psi_i$, resulting in an 
effective Majorana mass of the type:

$$M_R^{ij} \,=\, {1\over M_P}\,\Gamma_{ab}^{ij}\,{\bar V}_a {\bar V}_b\,.\eqno(34)$$

\noindent 
Recalling that, having set to zero the $F$ and the $D$ couplings 
in the superpotential (7,8), the standard neutrinos and 
anti-neutrinos cannot mix with the three (heavy) uneaten 
linear combinations of 
$\,\nu^c_{H_1}\,,\,\,\,\nu^c_{H_2}\,,\,\,\,{\bar\nu}^c_{H_1}\,$ 
and ${\bar\nu}^c_{H_2}\,$ we can write 
the neutrino mass matrix simply in the standard ($\nu_i,\nu^c_j$) 
basis\note{The situation is quite different in the previous 
minimal-type of $SU(5)$-flipped models, where the $F$ and the 
$D$ terms were present.}:

$${\cal M}_\nu \, =\, \left(\matrix{0&M_u\cr M_u^T&M_R\cr}\right)\,.\eqno(35)$$

\noindent 
This is just the ordinary seesaw mass matrix which,
in the general case where $M_R$ is non-singular, results in
three light neutrinos with a mass of order:

$$m_{\nu_i}\,\sim\, {m_{u_i}^2\over M_G^2}\,M_P\,.\eqno(36)$$

\noindent 
This formula will lead to a phenomenologically interesting 
neutrino mass spectrum:  
$m_{\nu_e} : m_{\nu_\mu} : m_{\nu_\tau} \simeq 
\, 10^{-9} : 10^{-3}-10^{-4} : 1-10 \,$ eV. For suitable 
choices of mixing angles, one may have MSW oscillations 
$\nu_e \rightarrow \nu_{\mu}\,$ in the sun [18] and we 
might ascribe the hot component of the Dark Matter of 
the universe to a $\tau$-neutrino with a mass of a few eV [19].

{\bf Discussion}

We have presented a viable version of the SO(10) 
model in which the breaking occurs via the flipped
rather than the usual Georgi-Glashow SU(5). The model
is consistent with the phenomenological requirements 
of having a non-trivial quark mixing matrix, natural 
doublet-triplet splitting, and a single pair of
light electroweak Higgs doublet scalar bosons. 
We recall that these requirements are in conflict
with the minimal version of the flipped models
embedded in SO(10). This conflict can not be 
solved by simply duplicating the doublet-triplet 
splitting mechanism characteristic of flipped 
models, through the use of a second pair of 
incomplete ${\bf 16} \oplus {\overline{\bf 16}}$ 
Higgs multiplets. We have remedied this situation
by adding instead a pair of complete spinorial
multiplets. 

We have also shown how, in the presence of suitable 
non-renormalizable superpotential terms, the model can 
reproduce the hierarchy observed 
in quark masses and mixings, as well as an acceptable
neutrino masses generated via the seesaw mechanism.

As a final comment we note that, so far in this paper we 
have assumed {\it ad hoc} the 
vanishing of the ``dangerous" $D,\,E,\,F$, and $L$ 
Yukawa-type coupling constants in the superpotential 
of eqs.(7,8). However, it is easy to see that it is possible 
to get rid of all these couplings by just introducing a $Z_2$ 
discrete symmetry in the superpotential, under which the 
$SO(10)$ superfields of the model transform as follows
\note{Actually, this is not the only possible choice; 
another solution, for example, is: $\Psi_i\rightarrow -\Psi_i\,,\,\,\,\,\,
\Delta_\alpha\rightarrow \Delta_\alpha\,,\,\,\,\,\,
\Theta_a\rightarrow \Theta_a\,,\,\,\,\,\,{\bar\Theta}_a\rightarrow -{\bar\Theta}_a\,,\,\,\,\,\,
\Phi\rightarrow -\Phi\,,\,\,\,\,\,\Sigma\rightarrow -\Sigma\,$.}:

$$\Psi_i\rightarrow \Psi_i\,,\,\,\,\,\,\Delta_\alpha\rightarrow \Delta_\alpha\,,\,\,\,\,\,
\Theta_a\rightarrow -\Theta_a\,,\,\,\,\,\,{\bar\Theta}_a\rightarrow {\bar\Theta}_a\,,\,\,\,\,\,
\Phi\rightarrow -\Phi\,,\,\,\,\,\,\Sigma\rightarrow -\Sigma\,.\eqno(37)$$ 

\noindent This symmetry also allows the presence of the 
non-renormalizable operators needed for implementing the 
neutrino seesaw mechanism, as discussed above. At a single 
generation level, this $Z_2$-symmetry is sufficient for the 
construction of a self-consistent model based on $SO(10)\rightarrow
SU(5)_{fl}\otimes U(1)$. 
Unfortunately, in a more realistic multi-generational scenario, our assumed
discrete symmetry is not sufficient to derive the particular 
structure of the non-renormalizable operators needed for reproducing
the correct charged fermion mass hierarchy in the framework of the
Georgi-Jarlskog {\sl texture}. In this case such a texture should 
follow from some underlying symmetry of the model. Moreover, the 
explanation of the masses of the first two generation 
charged leptons will require some 
extension of our scheme. Another difficulty of our
present model, is related to the $\mu$ problem [17]. Here we 
have ensured the absence of a dangerously large $\mu$-term, but in a 
more complete model one should be able to derive the correct value 
$\mu \sim M_W$.

\vfill 

We thank Ara Ioannissyan and Mario Gomez 
for helpful discussions. This work was supported by 
DGICYT under grant number PB92-0084 and by 
a postdoctoral fellowship from the European Union,
(S. R.). ERBCHBICT930726.

\vskip 3 cm

\noindent {\bf References}

\item{[1]} I. Antoniadis, J. Ellis, J.S. Hagelin and D.V. Nanopoulos, 
 Phys. Lett. {\bf 194B} (1987) 231;  {\bf 205B} (1988) 459;  {\bf 208B} (1988) 
209; {\bf 231B} (1989) 65.

\item{[2]}  I. Antoniadis, G.K. Leontaris and J. Rizos, Phys.
Lett. {\bf 245B} (1990) 161. 

\item{   } G.K. Leontaris, J. Rizos and K. Tamvakis,  Phys. 
Lett. {\bf 243B} (1990) 220; {\bf 251B} (1990) 83;

\item{   } I. Antoniadis, J. Rizos and K. Tamvakis, Phys. Lett. {\bf 278B}
(1992) 257; {\bf 279B} (1992) 281;

\item{   } J.L. Lopez and D.V. Nanopoulos, Nucl. Phys. {\bf B338} (1990) 73;
Phys. Lett. {\bf 251B} (1990) 73.

\item{   } D. Bailin and A. Love, Phys. Lett. {\bf 280B} (1992) 26.

\item{[3]} For a review see, {\it e.g.}, G. G. Ross, Grand-Unified
Theories, Benjamin, 1985; R. N. Mohapatra, Unification and 
Supersymmetry, Springer, 1986

\item{[4]} S. Dimopoulos, S. Wilczek, report N. NSF-ITP-82-07,
Aug. 1981 (unpublished); R. Cahn, I. Hinchliffe, L. Hall, 
Phys. Lett. {\bf 109B} (1982) 426.

\item{[5]} K. S. Babu,  R. N. Mohapatra, Phys. Rev. Lett. {\bf 74} (1995) 2418.

\item{[6]} S. M. Barr, Phys. Lett. {\bf 112B} (1982) 219.

\item{[7]} E. Papageorgiu and S. Ranfone, Phys. Lett. {\bf 282B} (1992) 89.

\item{[8]} H. Georgi, C. Jarlskog, Phys. Lett. {\bf 86B} (1979) 297.

\item{[9]} S. Ranfone and E. Papageorgiu, Phys. Lett. {\bf 295B} (1992) 79.

\item{ } S. Ranfone, Phys. Lett. {\bf 324B} (1994) 370.

\item{[10]} M. Martinez, talk at {\sl Elementary particle
Physics: Present and Future}, Ed. A. Ferrer and J. W. F. Valle,
World Scientific, in press.

\item{[11]}  Particle Data Group, Phys. Rev. {\bf D50} (1994) 1173.

\item{[12]} J. W. F. Valle, talk at TAUP 95, Toledo, ed. A. Morales 
{\it et al.}, Nucl. Phys. Proc. Suppl. (in press); A. Yu. Smirnov,
talk at {\sl Elementary particle Physics: Present and Future}, 
Ed. A. Ferrer and J. W. F. Valle, World Scientific, in press.

\item{[13]} 
 M. Gell-Mann, P. Ramond, R. Slansky,
in {\sl Supergravity},  ed. D. Freedman et al. (1979);
 T. Yanagida,
 in {\sl KEK lectures},  ed.  O. Sawada et al. (1979).

\item{[14]} E. Witten, Phys. Lett. {\bf B91} (1980) 81.

\item{[15] } 
E. Witten, Nucl. Phys. {\bf B258} (1985) 75;
R. Mohapatra, J. W. F. Valle, Phys. Rev. {\bf D34} (1986) 1642;
J. W. F. Valle, Nucl. Phys. Proc. Suppl. {\bf B11} (1989) 118;
see also refs. [1,2,7,9] above.

\item{[16] } J-P. Derendinger, L. Ibanez and H. P. Nilles,
Nucl. Phys. {\bf B267} (1986) 365; F. del Aguila et al.
Nucl. Phys. {\bf B272} (1986) 413; S. Nandi, U. Sarkar,
Phys. Rev. Lett. {\bf 55} (1986) 566; 
J. W. F. Valle, Phys. Lett. {\bf 186} (1987) 78

\item{[17] } 
C. Munoz, Proceedings of SUSY 94, Ann Arbor, USA,
C. Kolda, J. Wells,  editors.

\item{[18] } 
S. Mikheyev and A. Smirnov,  Sov. J. Nucl. 
Phys., {\bf 42} (1986) 1441; L. Wolfenstein, Phys. Rev., 
{\bf D17} (1978) 2369; {\bf D20}  (1979) 2634. 

\item{[19] } 
G.~F. Smoot et~al., Astr. J. {\bf 396} (1992) L1;
E.L.~Wright et al., Astr. J. {\bf 396} (1992) L13;
R. Rowan-Robinson, proceedings of the {\sl International
School on Cosmological Dark Matter}, 
Ed. A. Perez and J. W. F. Valle, World Scientific, 1994,
p. 7-13

\item{[[20] ] } 
N. Sakai, T. Yanagida, Nucl. Phys. {\bf B197} (1982) 533; 
S. Weinberg, Phys. Rev. {\bf D26} (1982) 287;
J. Ellis, D. Nanopoulos, S. Rudaz, Nucl. Phys. {\bf B202} (1982) 46;
S. Dimopoulos, S. Raby, F. Wilczek, Phys. Lett. {\bf B112} (1982) 133

\vfill\eject 
\bye